\documentclass[%
reprint,
showpacs,preprintnumbers,
amsmath,amssymb,
aps,
prb,
]{revtex4-1}

\usepackage{graphicx}
\usepackage{dcolumn}
\usepackage{bm}
\usepackage{colortbl}

\begin{document}


\title{Control of the non-stationary spin-polarized tunneling currents by applied bias changing}

\author{N.\,S.\,Maslova$^{1}$}
\altaffiliation{}
\author{P.\,I.\,Arseyev$^{2}$}
\author{V.\,N.\,Mantsevich$^{1}$}
\altaffiliation{} \email{vmantsev@gmail.com}

\affiliation{%
$^{1}$Moscow State University, 119991 Moscow, Russia, $^{2}$ P.N.
Lebedev Physical Institute RAS, 119991 Moscow, Russia
}%

\date{\today }
\begin{abstract}
We reveal that for the single Anderson impurity localized between
non-magnetic leads of the tunneling contact $"$magnetic$"$ state can
be distinguished from the $"$paramagnetic$"$ one only by the
analysis of the non-stationary system characteristics or the
behavior of the second order correlation functions for the localized
electrons occupation numbers. We investigate the response of the
system to the sudden shift of the applied bias and to the switching
$"$on$"$ the coupling to the second lead of the tunneling contact.
We demonstrate that in addition to the changes of the relaxation
regimes and typical relaxation time scales, non-stationary
spin-polarized currents flowing in the both leads are present in the
system. Spin polarization and direction of the non-stationary
currents in each lead can be simultaneously inverted by the sudden
changing of the applied bias voltage.
\end{abstract}

 \maketitle

\section{Introduction}
Physics of spin-polarized electron transport in semiconductor
nanostructures is among the most rapidly developing topics now a
days \cite{Awschalom}. Significant progress has been achieved in
experimental and theoretical investigation of spin-polarized
stationary transport in magnetic tunneling junctions \cite{Tsymbal},
\cite{Zutic}, \cite{Zhu}, \cite{Ohno}, \cite{Fiederling}. Different
magnetic materials, such as ferromagnetic metals \cite{Heersche} or
diluted magnetic semiconductors \cite{Egues} have been applied as a
spin injection sources and drains. Nevertheless spin-polarized
current sources, that use nonmagnetic materials are attractable as
they enable to avoid the presence of accidental magnetic field that
may cause undesirable effects on the spin currents. It was
demonstrated recently, that electron tunneling could be spin
dependent even in the case of nonmagnetic leads \cite{Perel'},
\cite{Glazov}. Moreover, spin-filter devices, which can generate a
spin-polarized current without using magnetic properties of
materials were proposed in \cite{Koga}, \cite{Voskoboynikov}.

To the best of our knowledge stationary spin-polarized currents are
usually under investigation. However, creation, diagnostics and
controlled manipulation of charge and spin states of the impurity
atoms or quantum dots (QDs), applicable for ultra small size
electronic devices design requires careful analysis of
non-stationary effects and transient processes
\cite{Bar-Joseph},\cite{Gurvitz_1},
\cite{Arseyev_1},\cite{Stafford_1},\cite{Hazelzet},\cite{Cota}.
Consequently, time dependent dynamics of initial spin and charge
configurations of correlated impurity or QD is an area of great
interest both from fundamental and technological point of view.
Moreover, characteristics of the stationary state of single impurity
interacting with the reservoir in the presence of strong Coulomb
correlations are not completely understood \cite{Arseyev_2},
\cite{Arseyev_3},
\cite{Contreras-Pulido},\cite{Elste},\cite{Kennes}.

In this paper we analyze non-stationary spin polarized currents
through the single-impurity state localized in the tunnel junction
in the presence of Coulomb correlations and applied bias voltage. We
demonstrate that $"$magnetic$"$ state can be distinguished from the
$"$paramagnetic$"$ one only by analyzing time evolution of opposite
spin electron occupation numbers. We reveal that non-stationary
spin-polarized currents can flow in the both leads and their
direction and polarization depend on the value of applied bias.
Moreover, non-stationary spin-polarized currents simultaneously
change their polarization and sign with the applied bias voltage
variations.

\section{Theoretical model}

We consider non-stationary processes in the system of single-level
impurity placed between two non-magnetic electronic reservoirs
(tunneling junction) with Coulomb correlations of localized
electrons. The Hamiltonian of the system

\begin{eqnarray}
\hat{H}=\hat{H}_{imp}+\hat{H}_{res}+\hat{H}_{tun}
\end{eqnarray}

is written as a sum of the single-level Anderson impurity
Hamiltonian

\begin{eqnarray}
\hat{H}_{imp}=\sum_{\sigma}\varepsilon_{1}\hat{n}_{1\sigma}++U\hat{n}_{1\sigma}\hat{n}_{1-\sigma},
\end{eqnarray}

non-magnetic electronic reservoirs Hamiltonian

\begin{eqnarray}
\hat{H}_{res}=\sum_{k\sigma}\varepsilon_{k}\hat{c}_{k\sigma}^{+}\hat{c}_{k\sigma}+\sum_{p\sigma}(\varepsilon_{p}-eV)\hat{c}_{p\sigma}^{+}\hat{c}_{p\sigma}
\end{eqnarray}

and the tunneling part

\begin{eqnarray}
\hat{H}_{tun}=
\sum_{k\sigma}t_{k}(\hat{c}_{k\sigma}^{+}\hat{c}_{1\sigma}+\hat{c}_{1\sigma}^{+}\hat{c}_{k\sigma})+
\sum_{p\sigma}t_{p}(\hat{c}_{p\sigma}^{+}\hat{c}_{1\sigma}+\hat{c}_{1\sigma}^{+}\hat{c}_{p\sigma}).\nonumber\\
\end{eqnarray}

Here index $k(p)$ labels continuous spectrum states in the leads,
$t_{k(p)}$ is the tunneling transfer amplitude between continuous
spectrum states and localized state with the energy $\varepsilon_1$
which is considered to be independent on the momentum and spin.
Operators $\hat{c}_{k(p)}^{+}/\hat{c}_{k(p)}$ are the
creation/annihilation operators for the electrons in the continuous
spectrum states $k(p)$.
$\hat{n}_{1\sigma(-\sigma)}=\hat{c}_{1\sigma(-\sigma)}^{+}\hat{c}_{1\sigma(-\sigma)}$-localized
state electron occupation numbers, where operator
$\hat{c}_{1\sigma(-\sigma)}$ destroys electron with the spin
$\sigma(-\sigma)$ on the energy level $\varepsilon_1$. $U$ is the
on-site Coulomb repulsion for the double occupation of the localized
state. Our analysis deals with the low temperature regime when the
Fermi level is well defined and the temperature is much lower than
all the typical energy scales in the system. Consequently the
distribution function of the electrons in the leads (band electrons)
is close to the Fermi step.

\section{Non-stationary electronic transport: formalism and results}

Let us further consider $\hbar=1$ and $e=1$ elsewhere, so the motion
equations for the electron operators products
$\hat{c}_{1\sigma}^{+}\hat{c}_{1\sigma}$,
$\hat{c}_{1\sigma}^{+}\hat{c}_{k(p)\sigma}$ and
$\hat{c}_{k^{'}(p)\sigma}^{+}\hat{c}_{k(p)\sigma}$ can be written
as:

\begin{eqnarray}
i\frac{\partial \hat{c}_{1\sigma}^{+}\hat{c}_{1\sigma}}{\partial
t}=&-&\sum_{k,\sigma}t_{k}(\hat{c}_{k\sigma}^{+}\hat{c}_{1\sigma}-\hat{c}_{1\sigma}^{+}\hat{c}_{k\sigma})-\nonumber\\&-&
\sum_{p,\sigma}t_{p}(\hat{c}_{p\sigma}^{+}\hat{c}_{1\sigma}-\hat{c}_{1\sigma}^{+}\hat{c}_{p\sigma}),
\label{1}
\end{eqnarray}

\begin{eqnarray}
i\frac{\partial \hat{c}_{1\sigma}^{+}\hat{c}_{k\sigma}}{\partial
t}=&-&(\varepsilon_1-\varepsilon_k)
\hat{c}_{1\sigma}^{+}\hat{c}_{k\sigma}-\nonumber\\&-&U\hat{n}_{1-\sigma}
\hat{c}_{1\sigma}^{+}\hat{c}_{k\sigma}
+t_{k}(\hat{n}_{1\sigma}-\hat{n}_{k\sigma})
-\nonumber\\&-&\sum_{k^{'}\neq
k}t_{k^{'}}\hat{c}_{k^{'}\sigma}^{+}\hat{c}_{k\sigma}-
\sum_{p}t_{p}\hat{c}_{p\sigma}^{+}\hat{c}_{k\sigma},\label{2}
\end{eqnarray}

\begin{eqnarray}
i\frac{\partial \hat{c}_{k^{'}\sigma}^{+}\hat{c}_{k\sigma}}{\partial
t}=&-&(\varepsilon_{k^{'}}-\varepsilon_k)\cdot
\hat{c}_{k^{'}\sigma}^{+}\hat{c}_{k\sigma}-\nonumber\\&-&t_{k^{'}}
\hat{c}_{1\sigma}^{+}\hat{c}_{k\sigma}+t_{k}
\hat{c}_{k^{'}\sigma}^{+}\hat{c}_{1\sigma} \label{3}
\end{eqnarray}

and

\begin{eqnarray}
i\frac{\partial \hat{c}_{p\sigma}^{+}\hat{c}_{k\sigma}}{\partial
t}=&(&\varepsilon_{k}-\varepsilon_p)\cdot
\hat{c}_{p\sigma}^{+}\hat{c}_{k\sigma}+\nonumber\\&+&t_{k}\cdot
\hat{c}_{p\sigma}^{+}\hat{c}_{1\sigma}-t_{p}\cdot
\hat{c}_{1\sigma}^{+}\hat{c}_{k\sigma},
\end{eqnarray}

where $\hat{n}_{k}=\hat{c}_{k\sigma}^{+}\hat{c}_{k\sigma}$ is an
occupation operator for the electrons in the reservoir.

Equations of motion for the electron operators products
$\hat{c}_{1\sigma}^{+}\hat{c}_{p\sigma}$ and
$\hat{c}_{p^{'}\sigma}^{+}\hat{c}_{p\sigma}$ can be obtained from
Eq.(\ref{2}) and Eq. (\ref{3}) correspondingly  by the indexes
substitution $k\leftrightarrow p$ and $k^{'}\leftrightarrow p^{'}$.

One can easily obtain:

\begin{eqnarray}
\sum_{k^{'}\neq
k}\hat{c}_{k^{'}\sigma}^{+}\hat{c}_{k\sigma}t_{k^{'}}&=&i\sum_{k^{'}}\int^{t}
dt_{1}e^{i\cdot(\varepsilon_{k}-\varepsilon_k^{'})\cdot(t-t_{1})}\times\nonumber\\&\times&(t_{k^{'}}^{2}\cdot
\hat{c}_{1\sigma}^{+}\hat{c}_{k\sigma}-t_{k}t_{k^{'}}\cdot
\hat{c}_{k^{'}\sigma}^{+}\hat{c}_{1\sigma})\label{4}.\nonumber\\
\end{eqnarray}

Similar expressions for the products
$\hat{c}_{p^{'}\sigma}^{+}\hat{c}_{p\sigma}$,
$\hat{c}_{p^{'}\sigma}^{+}\hat{c}_{k\sigma}$ and so on can be
obtained from expression (\ref{4}) by the indexes changing
$k\leftrightarrow p$,$k^{'}\leftrightarrow p^{'}$ and so on.

Substituting (\ref{4}) in Eq. (\ref{2}) we get

\begin{eqnarray}
i\frac{\partial \hat{c}_{1\sigma}^{+}\hat{c}_{k\sigma}}{\partial
t}&+&(\varepsilon_1-\varepsilon_k+i\Gamma)\hat{c}_{1\sigma}^{+}\hat{c}_{k\sigma}+\nonumber\\&+&U\hat{n}_{1-\sigma}
\hat{c}_{1\sigma}^{+}\hat{c}_{k\sigma}=t_{k}(\hat{n}_{1\sigma}-\hat{n}_{k\sigma})+\nonumber\\&+&i\sum_{k^{'}}\int^{t}
dt_{1}t_{k}t_{k^{'}}\hat{c}_{k^{'}\sigma}^{+}(t_{1})\hat{c}_{1\sigma}(t_{1})
e^{i\cdot(\varepsilon_{k}-\varepsilon_k^{'})\cdot(t-t_{1})}
+\nonumber\\&+&i\sum_{p}\int^{t}
dt_{1}t_{k}t_{p}\hat{c}_{p\sigma}^{+}(t_{1})\hat{c}_{1\sigma}(t_{1})e^{i\cdot(\varepsilon_{k}-\varepsilon_p)\cdot(t-t_{1})},\nonumber\\
\label{5}
\end{eqnarray}

where $\Gamma=\Gamma_k+\Gamma_p$ and
$\Gamma_{k(p)}=\nu_{k(p)0}t_{k(p)}^{2}$, $\nu_{k(p)0}$ - are the
unperturbed densities of states in the left and right leads of the
tunneling contact.

Equation for $\frac{\partial
\hat{c}_{1\sigma}^{+}\hat{c}_{p\sigma}}{\partial t}$ can be obtained
from Eq.(\ref{5}) by the indexes $k\leftrightarrow p$ and
$k^{'}\leftrightarrow p^{'}$ changing. Multiplying Eq. (\ref{5}) by
the electron operators $1-\hat{n}_{1-\sigma}$ and
$\hat{n}_{1-\sigma}$ we obtain the following expressions:

\begin{eqnarray}
(1-\hat{n}_{1-\sigma})&\cdot& i\frac{\partial
\hat{c}_{1\sigma}^{+}\hat{c}_{k\sigma}}{\partial
t}+(\varepsilon_1-\varepsilon_k+i\Gamma)(1-\hat{n}_{1-\sigma})\hat{c}_{1\sigma}^{+}\hat{c}_{k\sigma}=\nonumber\\&=&(1-\hat{n}_{1-\sigma})\cdot
[t_{k}\cdot(\hat{n}_{1\sigma}-\hat{n}_{k\sigma})+\nonumber\\&+&i\sum_{k^{'}}\int^{t}
dt_{1}t_{k}t_{k^{'}}\hat{c}_{k^{'}\sigma}^{+}(t_{1})\hat{c}_{1\sigma}(t_{1})e^{i\cdot(\varepsilon_{k}-\varepsilon_k^{'})\cdot(t-t_{1})}+\nonumber\\&+&i\sum_{p}\int^{t}
dt_{1}t_{k}t_{p}\hat{c}_{p\sigma}^{+}(t_{1})\hat{c}_{1\sigma}(t_{1})e^{i\cdot(\varepsilon_{k}-\varepsilon_p)\cdot(t-t_{1})}
]\nonumber\\
\end{eqnarray}

and

\begin{eqnarray}
\hat{n}_{1-\sigma}&\cdot& i\frac{\partial
\hat{c}_{1\sigma}^{+}\hat{c}_{k\sigma}}{\partial
t}+(\varepsilon_1-\varepsilon_k+U+i\Gamma)\hat{n}_{1-\sigma}\hat{c}_{1\sigma}^{+}\hat{c}_{k\sigma}=\nonumber\\&=&\hat{n}_{1-\sigma}\cdot
[t_{k}\cdot(\hat{n}_{1\sigma}-\hat{n}_{k\sigma})+\nonumber\\&+&i\sum_{k^{'}}\int^{t}
dt_{1}t_{k}t_{k^{'}}\hat{c}_{k^{'}\sigma}^{+}(t_{1})\hat{c}_{1\sigma}(t_{1})e^{i\cdot(\varepsilon_{k}-\varepsilon_k^{'})\cdot(t-t_{1})}+\nonumber\\&+&i\sum_{p}\int^{t}
dt_{1}t_{k}t_{p}\hat{c}_{p\sigma}^{+}(t_{1})\hat{c}_{1\sigma}(t_{1})e^{i\cdot(\varepsilon_{k}-\varepsilon_p)\cdot(t-t_{1})}
].\nonumber\\
\label{6}
\end{eqnarray}

If condition $\frac{\varepsilon_1-\varepsilon_F}{\Gamma}>>1$ is
fulfilled, $\hat{n}_{1-\sigma}$ is a slowly varying quantity in
comparison with the $\hat{c}_{1\sigma}^{+}\hat{c}_{k(p)\sigma}$
($\frac{\partial}{\partial
t}\hat{n}_{1-\sigma}<<\frac{\partial}{\partial
t}\hat{c}_{1\sigma}^{+}\hat{c}_{k(p)\sigma}$). Consequently, it is
reasonable to consider that:

\begin{eqnarray}
\frac{\partial}{\partial
t}\hat{n}_{1-\sigma}\hat{c}_{1\sigma}^{+}\hat{c}_{k(p)\sigma}\sim
\hat{n}_{1-\sigma}\frac{\partial}{\partial
t}\hat{c}_{1\sigma}^{+}\hat{c}_{k(p)\sigma}.
\label{42}\end{eqnarray}

So, the terms $(\frac{\partial}{\partial
t}\hat{n}_{1-\sigma})\hat{c}_{1\sigma}^{+}\hat{c}_{k(p)\sigma}$  are
omitted. One can get expressions for
$(1-\widehat{n}_{1-\sigma})\hat{c}_{1\sigma}^{+}\hat{c}_{k(p)\sigma}$
and
$\widehat{n}_{1-\sigma}\hat{c}_{1\sigma}^{+}\hat{c}_{k(p)\sigma}$
[using the procedure similar to the one which was used to obtain
Eq.(\ref{4}) from Eq.(\ref{3})] and then for the
$\hat{c}_{1\sigma}^{+}\hat{c}_{k(p)\sigma}$.

Substituting expressions for the
$\hat{c}_{1\sigma}^{+}\hat{c}_{k(p)\sigma}$ and
$\hat{c}_{k(p)\sigma}^{+}\hat{c}_{1\sigma}$ to Eq. (\ref{1}) we
obtain equations, which determine time evolution of the electron
occupation numbers $\hat{n}_{1\sigma}$. It is necessary to note,
that the last term in Eq. (\ref{5}) after summation over the index
$k(p)$ doesn't contribute to the non-stationary equations for the
electron occupation numbers $\hat{n}_{1\sigma}$. So, time evolution
of the electron occupation numbers operators in the situation when
the second lead and the non-zero bias voltage are present can be
analyzed by means of the system of equations:

\begin{eqnarray}
\frac{\hat{n}_{1\sigma}}{\partial
t}&=&-2\Gamma[\hat{n}_{1\sigma}-(1-\hat{n}_{1-\sigma})
\hat{N}_{\varepsilon}^{T}(t)-\hat{n}_{1-\sigma}
\hat{N}_{\varepsilon+U}^{T}(t)],\nonumber\\
\frac{\hat{n}_{1-\sigma}}{\partial
t}&=&-2\Gamma[\hat{n}_{1-\sigma}-(1-\hat{n}_{1\sigma})
\hat{N}_{\varepsilon}^{T}(t)-\hat{n}_{1\sigma}
\hat{N}_{\varepsilon+U}^{T}(t)], \label{8}\nonumber\\
\end{eqnarray}

where

\begin{eqnarray}
\hat{N}_{\varepsilon}^{T}(t)&=&\frac{\Gamma_k}{\Gamma}\cdot\hat{N}_{k\varepsilon}(t)+\frac{\Gamma_p}{\Gamma}\cdot\hat{N}_{p\varepsilon}(t),\nonumber\\
\hat{N}_{\varepsilon+U}^{T}(t)&=&\frac{\Gamma_k}{\Gamma}\cdot\hat{N}_{k\varepsilon+U}(t)+\frac{\Gamma_p}{\Gamma}\cdot\hat{N}_{p\varepsilon+U}(t)\nonumber\\
\label{401}\end{eqnarray}

and

\begin{eqnarray}
\hat{N}_{k(p)\varepsilon}=\frac{1}{2}i\cdot \int
d\varepsilon_{k(p)}f_{k(p)}^{\sigma}(\varepsilon_{k(p)})\times\nonumber\\\times[\frac{1-e^{i(\varepsilon_1+i\widetilde{\Gamma}-\varepsilon_{k(p)})t}}{\varepsilon_1+i\widetilde{\Gamma}-\varepsilon_{k(p)}}-\frac{1-e^{-i(\varepsilon_1-i\widetilde{\Gamma}-\varepsilon_{k(p)})t}}{\varepsilon_1-i\widetilde{\Gamma}-\varepsilon_{k(p)}}],\nonumber\\
\hat{N}_{k(p)\varepsilon+U}(t)=\frac{1}{2}i\cdot \int
d\varepsilon_{k(p)}f_{k(p)}^{\sigma}(\varepsilon_{k(p)})\times\nonumber\\\times[\frac{1-e^{i(\varepsilon_1+U+i\widetilde{\Gamma}-\varepsilon_{k(p)})t}}{\varepsilon_1+U+i\widetilde{\Gamma}-\varepsilon_{k(p)}}-\frac{1-e^{-i(\varepsilon_1+U-i\widetilde{\Gamma}-\varepsilon_{k(p)})t}}{\varepsilon_1+U-i\widetilde{\Gamma}-\varepsilon_{k(p)}}].\nonumber\\
\label{41}\end{eqnarray}

In Eq.(\ref{41}) parameter $\Gamma$ is replaced by the effective
parameter $\widetilde{\Gamma}\simeq\Gamma$, because after applying
approximation (\ref{42}) to Eq. (\ref{6}) the omitted terms are of
the order of $\Gamma$. We can obtain equations for the occupation
numbers of localized electrons $n_{1\pm\sigma}$  by averaging Eqs.
(\ref{8})-(\ref{41}) for the operators and by decoupling electrons
occupation numbers in the reservoir. Such decoupling procedure is
reasonable if one considers that electrons in the macroscopic
reservoir are in the thermal equilibrium. After decoupling one has
to replace electron occupation numbers operators in the reservoir
$\hat{n}_{k}^{\sigma}$ in Eqs. (\ref{401})-(\ref{41}) by the Fermi
distribution functions $f_{k}^{\sigma}$.

If one is interested in the situation when the second lead and
non-zero bias voltage are switched $"$on$"$ at the time moment
$t=t_0>0$, Eqs. (\ref{8}) can be easily generalized:

\begin{eqnarray}
\frac{\partial n_{1\sigma}}{\partial
t}=-2\Theta(t_{0}-t)\Gamma\times\nonumber\\
\times[n_{1\sigma}-(1-n_{1-\sigma}) N_{k\varepsilon}(t)-n_{1-\sigma}
N_{k\varepsilon+U}(t)]-\nonumber\\
-2\Theta(t-t_{0})\Gamma\times\nonumber\\
\times[n_{1\sigma}-(1-n_{1-\sigma})
N_{k\varepsilon}^{T}(t)-n_{1-\sigma}
N_{k\varepsilon+U}^{T}(t)],\nonumber\\
\frac{\partial n_{1-\sigma}}{\partial
t}=-2\Theta(t_{0}-t)\Gamma\times\nonumber\\
\times[n_{1-\sigma}-(1-n_{1\sigma}) N_{k\varepsilon}(t)-n_{1\sigma}
N_{k\varepsilon+U}(t)]-\nonumber\\
-2\Theta(t-t_{0})\Gamma\times\nonumber\\
\times[n_{1-\sigma}-(1-n_{1\sigma})
N_{k\varepsilon}^{T}(t)-n_{1\sigma}
N_{k\varepsilon+U}^{T}(t)],\nonumber\\
\end{eqnarray}

where $\Gamma=\Gamma_k$ for $t<t_0$ and $\Gamma=\Gamma_k+\Gamma_p$
for $t>t_0$. Solution can be easily obtained by the numerical
simulations of the system of equations. Time dependent dynamics of
the electron occupation numbers and their correlation functions can
be analyzed for the different initial conditions: 1) the non-zero
localized magnetic moment exists on the impurity
($|n_{1\sigma}-n_{1-\sigma}|\sim1$). Such state can be prepared due
to the applied external magnetic field $\mu B>>\varepsilon_1$, which
is switched $"$off$"$ at the initial time moment $t=0$; 2) the
initial state close to the highly occupied paramagnetic one
($|1-n_{1\pm\sigma}|<<1$) can be prepared by the applied bias
voltage $|eV|>\varepsilon_1+U$ switching $"$off$"$ or $"$on$"$ at
the initial time moment $t=0$; 3) the initial state close to the low
occupied paramagnetic one ($|n_{1\pm\sigma}|<<1$) can be prepared by
the applied bias voltage $|eV|<\varepsilon_1$ switching $"$off$"$ or
$"$on$"$ at the initial time moment $t=0$. It will be shown that
relaxation time scale strongly depends on the properties of the
initially prepared state.

If one is interested in system time evolution for the time scales
$t>>\frac{1}{\varepsilon_1}$, fast oscillating terms, which contain
time dependent exponents can be neglected and functions
$N_{k(p)\varepsilon}$, $N_{k(p)\varepsilon+U}$ [see Eq.(\ref{41})]
become independent from $t$. So, the localized electrons occupation
numbers $n_{1\sigma}(t)$, $n_{1-\sigma}(t)$  can be easily found for
the arbitrary initial conditions:

For $0<t<t_0$

\begin{eqnarray}
n_{1\sigma}(t)=\frac{N_{k\varepsilon}}{1+\Delta
N}\cdot(1-e^{\lambda_2t})+\nonumber\\+\frac{n_{1\sigma}(0)-n_{1-\sigma}(0)}{2}\cdot
e^{\lambda_1t}+\frac{n_{1\sigma}(0)+n_{1-\sigma}(0)}{2}\cdot
e^{\lambda_2t},\nonumber\\
n_{1-\sigma}(t)=\frac{N_{k\varepsilon}}{1+\Delta
N}\cdot(1-e^{\lambda_2t})+\nonumber\\+\frac{n_{1-\sigma}(0)-n_{1\sigma}(0)}{2}\cdot
e^{\lambda_1t}+\frac{n_{1-\sigma}(0)+n_{1\sigma}(0)}{2}\cdot
e^{\lambda_2t},\nonumber\\
\label{43}\end{eqnarray}

where $n_{1\pm\sigma}(0)$ are the initial conditions. For $t>t_0$

\begin{eqnarray}
n_{1\sigma}(t)&=&\frac{N_{\varepsilon}^{T}}{1+\Delta
N^{T}}\cdot(1-e^{\lambda_{2}^{T}(t-t_{0})})+\nonumber\\&+&\frac{n_{1\sigma}(t_0)-n_{1-\sigma}(t_0)}{2}\cdot
e^{\lambda_{1}^{T}(t-t_{0})}+\nonumber\\&+&\frac{n_{1\sigma}(t_0)+n_{1-\sigma}(t_0)}{2}\cdot
e^{\lambda_{2}^{T}(t-t_{0})},\nonumber\\
n_{1-\sigma}(t)&=&\frac{N_{k\varepsilon}^{T}}{1+\Delta
N^{T}}\cdot(1-e^{\lambda_{2}^{T}(t-t_{0})})+\nonumber\\&+&\frac{n_{1-\sigma}(t_0)-n_{1\sigma}(t_0)}{2}\cdot
e^{\lambda_{1}^{T}(t-t_{0})}+\nonumber\\&+&\frac{n_{1-\sigma}(t_0)+n_{1\sigma}(t_0)}{2}\cdot
e^{\lambda_{2}^{T}(t-t_{0})}.\nonumber\\
\end{eqnarray}

Eigenvalues $\lambda_{1,2}$ and $\lambda_{1,2}^{T}$ have the
following form:

\begin{eqnarray}
\lambda_{1,2}&=&-2\Gamma\cdot(1\mp\Delta N),\nonumber\\
\lambda_{1,2}^{T}&=&-2\Gamma\cdot(1\mp\Delta N^{T})
\end{eqnarray}

and

\begin{eqnarray}
\Delta N&=&N_{\varepsilon}-N_{\varepsilon+U},\nonumber\\
\Delta N^{T}&=&N_{\varepsilon}^{T}-N_{\varepsilon+U}^{T}.
\end{eqnarray}

\begin{figure}
\includegraphics[width=70mm]{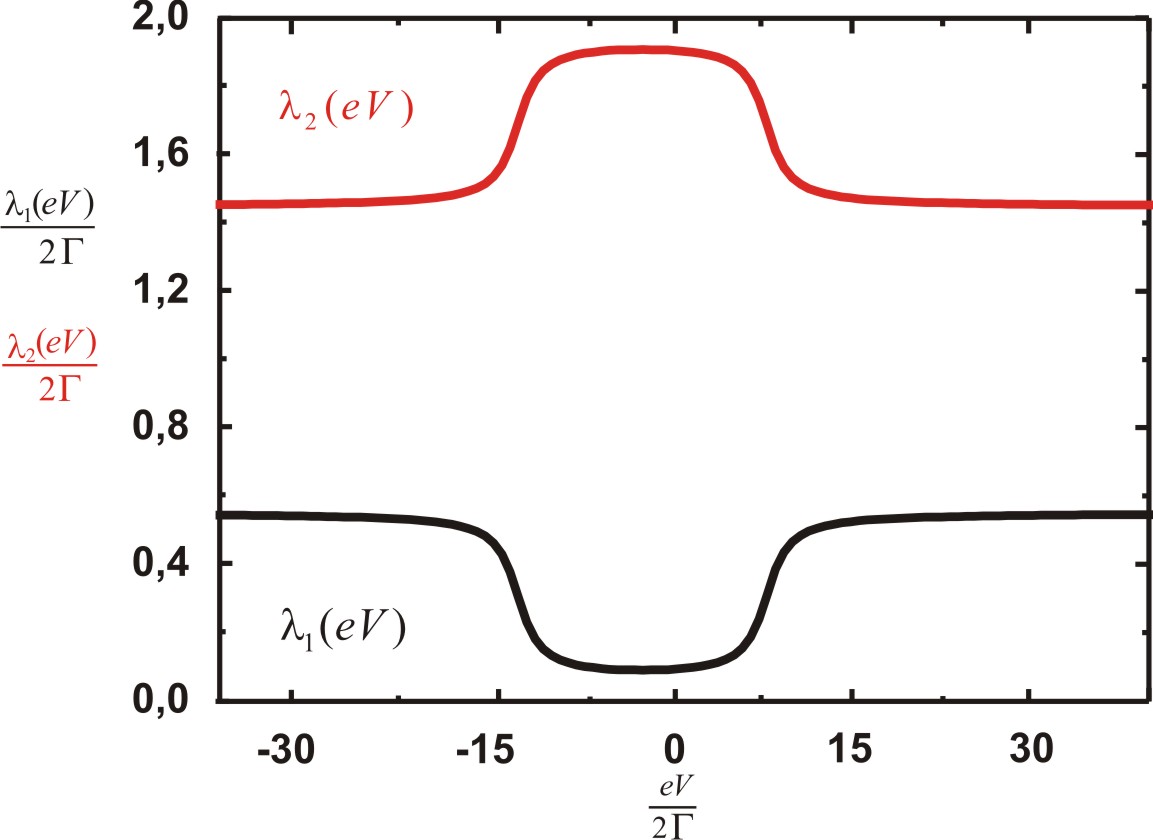}%
\caption{(Color online) Normalized relaxation rates behavior as a
functions of the applied bias voltage for the parameters
$\varepsilon_1/2\Gamma=-2.5$ $U/2\Gamma=7.5$ and
$\Gamma=\Gamma_k+\Gamma_p=1$ ($\Gamma_k=\Gamma_p$). }
\label{figure1}
\end{figure}

$n_{1\pm\sigma}(t_0)$ is determined by  expressions (\ref{43}) for
$t=t_0$. Relaxation rates behavior as a function of applied bias
voltage is shown in the Fig.\ref{figure1}. For small values of
applied bias $eV<\varepsilon$ two time scales
$|\lambda_{1,2}^{T}|^{-1}$ strongly differ, while for the large
absolute values of the applied bias
$\frac{|\lambda_{1}^{T}|}{|\lambda_{2}^{T}|}\rightarrow1/3$.

For the infinitely large times $t\rightarrow\infty$ the stationary
state is always $"$paramagnetic$"$ one and electron occupation
numbers are:

\begin{eqnarray}
n_{1\sigma}^{stT}=n_{1-\sigma}^{stT}=\frac{N_{\varepsilon}^{T}}{1+\Delta
N^{T}}
\end{eqnarray}

\begin{figure}[h!]
\includegraphics[width=80mm]{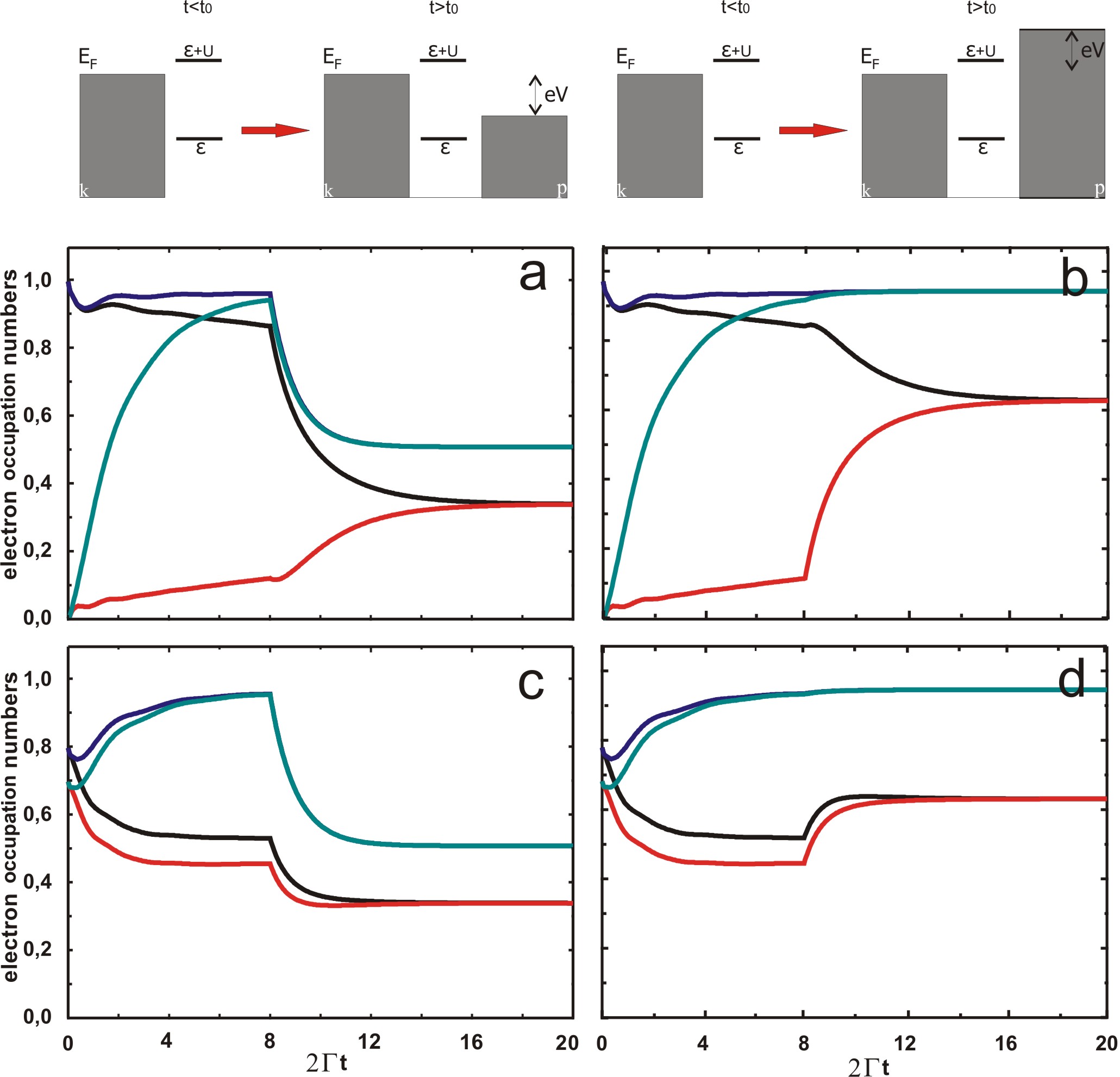}%
\caption{(Color online) Changing of relaxation regimes of the
electron occupation numbers when coupling to the second lead is
switched $"$on$"$ at $2\Gamma t_{0}=8$ for different initial
conditions. Black line demonstrates $n_{1\sigma}(t)$ and red line -
$n_{1-\sigma}(t)$ in the presence of Coulomb correlations. Blue line
demonstrates $n_{1\sigma}(t)$ and green line - $n_{1-\sigma}(t)$ in
the absence of Coulomb correlations. a),b) $n_{1\sigma}=1$,
$n_{1-\sigma}=0$; c),d) $n_{1\sigma}=0.8$, $n_{1-\sigma}=0.7$. a),c)
$eV/2\Gamma=-5.0$ and b),d) $eV/2\Gamma=12.5$.  Parameters
$U/2\Gamma=7.5$, $\varepsilon/2\Gamma=-2.5$, and $\Gamma=1$
($\Gamma_k=\Gamma_p$) are the same for all the figures.}
\label{figure2}
\end{figure}

The behavior of localized state electron occupation numbers for the
different initial conditions and the set of system parameters in the
case, when the second lead is switched on at the time moment
$t=t_0>0$ is depicted in the Fig.(\ref{figure2}). Obtained results
demonstrate, that switching $"$on$"$ of the second lead with the
non-zero applied bias results in the increasing of the relaxation
rate and consequently destroys the long-living $"$magnetic$"$
moment.

\section{Non-stationary spin-polarized currents: formalism and results}

If the initial state is a $"$magnetic$"$ one, non-stationary
spin-polarized currents $I_{k(p)}(t)^{\pm}$ flow in the each lead:

\begin{eqnarray}
I_{k}^{\pm}(t)&=&-2\Gamma_{k}[n_{1\pm\sigma}-(1-n_{1\mp\sigma})
N_{k\varepsilon}(t)-n_{1\mp\sigma}
N_{k\varepsilon+U}(t)],\nonumber\\
I_{p}^{\pm}(t)&=&-2\Gamma_{p}[n_{1\pm\sigma}-(1-n_{1\mp\sigma})
N_{p\varepsilon}(t)-n_{1\mp\sigma}
N_{p\varepsilon+U}(t)],\nonumber\\
\end{eqnarray}

where electron occupation numbers $n_{1\pm\sigma}$ are determined
from the system of equations (\ref{8}) with the magnetic initial
conditions.

\begin{figure}
\includegraphics[width=70mm]{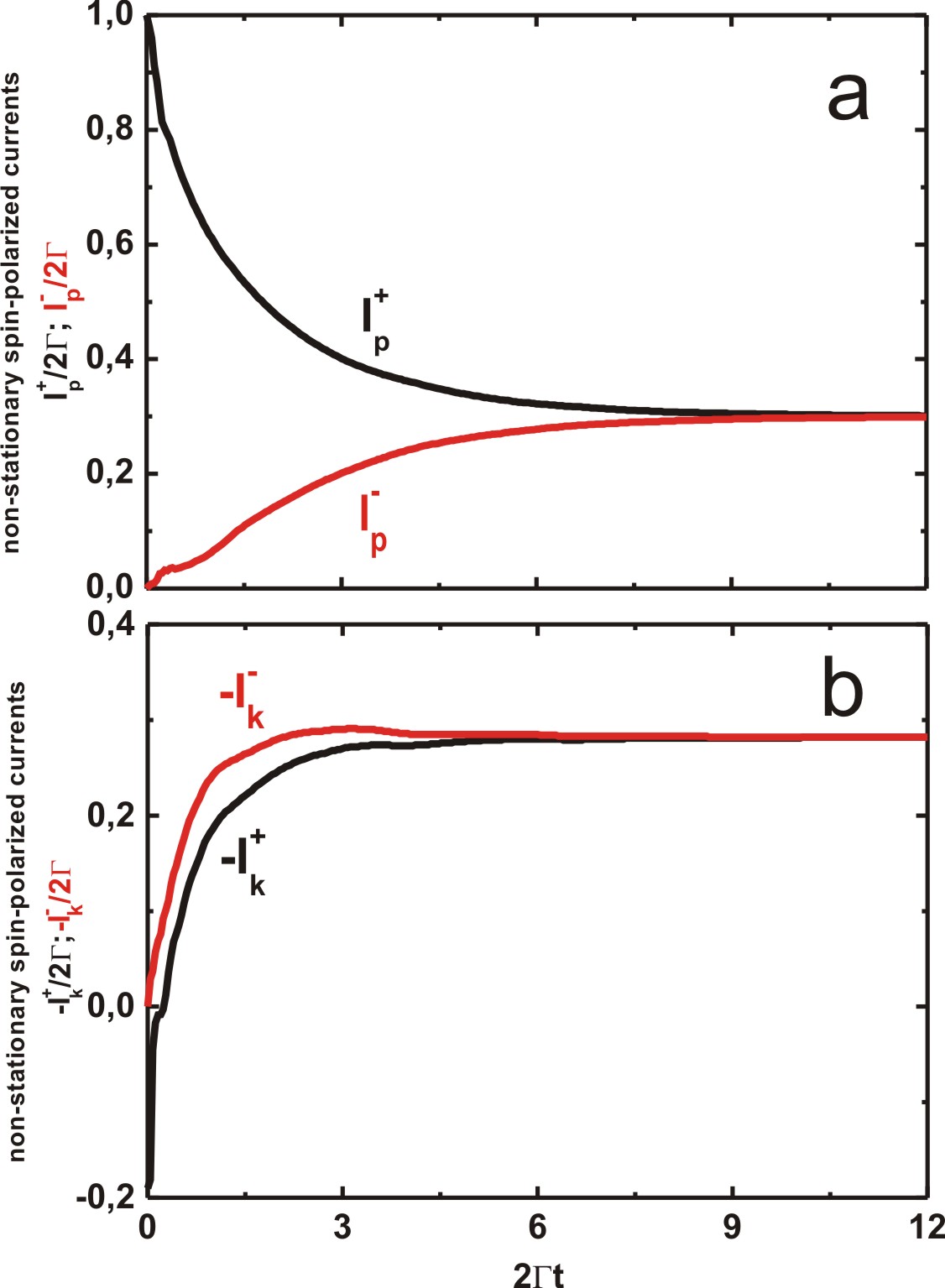}%
\caption{(Color online) Normalized non-stationary spin-polarized
tunneling currents $I_{k(p)}^{+}(t)/2\Gamma$ (black line) and
$I_{k(p)}^{-}(t)/2\Gamma$ (red line). a) Black line $I_{p}^{+}(t)$,
red line $I_{p}^{-}(t)$; b) Black line $I_{k}^{+}(t)$, red line
$I_{k}^{-}(t)$. Parameters $U/2\Gamma=7.5$,
$\varepsilon/2\Gamma=-2.5$, $eV/2\Gamma=-5.0$ and $\Gamma=1$ are the
same for all the figures. $n_{1\sigma}(0)=1$, $n_{1-\sigma}(0)=0$.}
\label{figure3}
\end{figure}

\begin{figure}
\includegraphics[width=70mm]{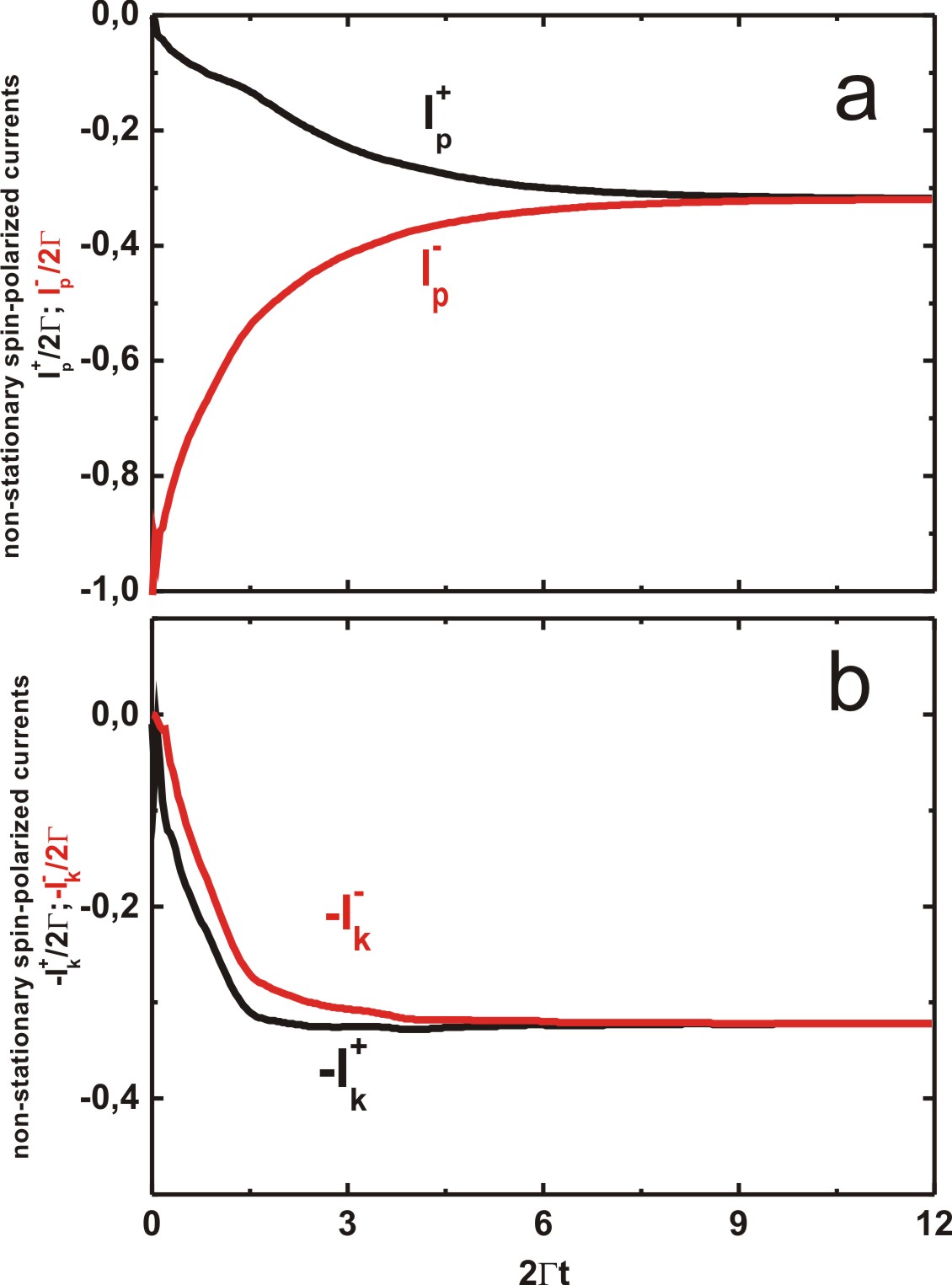}%
\caption{(Color online) Normalized non-stationary spin-polarized
tunneling currents $I_{k(p)}^{+}(t)/2\Gamma$ (black line) and
$I_{k(p)}^{-}(t)/2\Gamma$ (red line). a) Black line $I_{p}^{+}(t)$,
red line $I_{p}^{-}(t)$; b) Black line $I_{k}^{+}(t)$, red line
$I_{k}^{-}(t)$. Parameters $U/2\Gamma=7.5$,
$\varepsilon/2\Gamma=-2.5$, $eV/2\Gamma=12.5$ and $\Gamma=1$ are the
same for all the figures. $n_{1\sigma}(0)=1$, $n_{1-\sigma}(0)=0$.}
\label{figure4}
\end{figure}

\begin{figure}
\includegraphics[width=70mm]{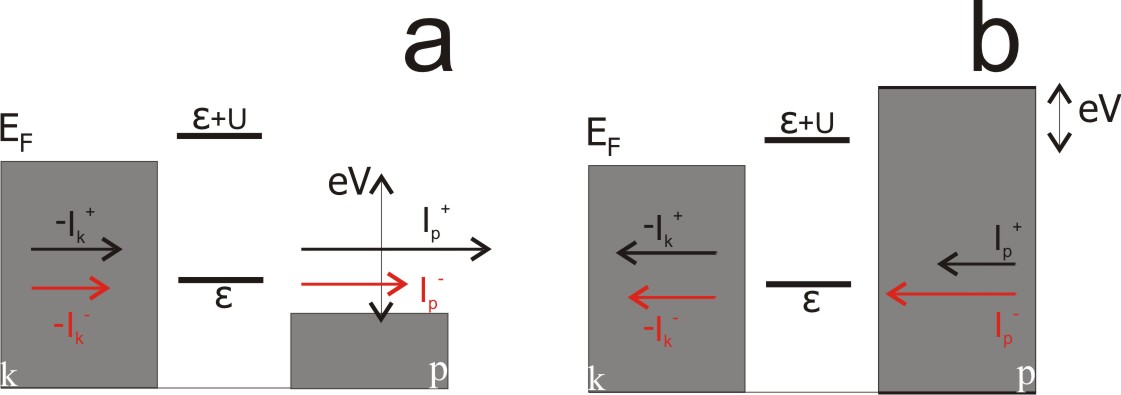}%
\caption{(Color online) Schemes of the spin-polarized currents
directions. Scheme a) corresponds to the results, shown in the
Fig.\ref{figure3}, b) in the Fig.\ref{figure4}. } \label{figure5}
\end{figure}

Non-stationary spin-polarized currents can flow in the both leads
and their direction and polarization depend on the value of applied
bias. Non-stationary spin-polarized currents $I_{k(p)}(t)^{\pm}$ for
the initially prepared $"$magnetic$"$ state for the different
constant values of applied bias are depicted in Fig.(\ref{figure3})-
Fig.(\ref{figure4}). Schemes of the spin-polarized currents
directions are shown in Fig.(\ref{figure5}). For the large negative
values of applied bias voltage (see Fig.\ref{figure3}a and
Fig.\ref{figure5}a) non-stationary spin-polarized  currents
$I_{p}^{+}(t)$ and $I_{p}^{-}(t)$ are flowing in the same direction
in the tunneling contact lead with the Fermi level shifted by the
applied bias voltage (lead $p$). In this case strong spin
polarization of the total current occurs at the initial stage of
relaxation as the amplitude of current $I_{p}^{+}(t)$ strongly
exceeds the amplitude of current $I_{p}^{-}(t)$. Non-stationary
spin-polarized currents $I_{k}^{+}(t)$ and $I_{k}^{-}(t)$ in the
lead with $E_F=0$ are also flowing in the same direction, but the
difference between currents amplitudes is small (see
Fig.\ref{figure3}b and Fig.\ref{figure5}a).

For the positive values of applied bias voltage (see
Fig.\ref{figure4} and Fig.\ref{figure5}b) direction of the
non-stationary spin-polarized currents changes to the opposite one
in comparison with the case, when large negative bias was applied to
the tunneling contact [see Fig.\ref{figure5}a]. One can easily
distinguish the presence of non-stationary spin-polarized  currents
$I_{p}^{+}(t)$ and $I_{p}^{-}(t)$ again flowing in the same
direction in the tunneling contact lead with the Fermi level shifted
by the applied bias voltage (lead $p$) (see Fig.\ref{figure4}a and
Fig.\ref{figure5}b). Non-stationary spin-polarized currents
$I_{k}^{+}(t)$ and $I_{k}^{-}(t)$ in the lead with $E_F=0$ are also
flowing in the same direction, but the difference between currents
amplitudes is quite small (see Fig.\ref{figure4}b and
Fig.\ref{figure5}b). Fig. (\ref{figure3})-(\ref{figure4})
demonstrate equal amplitudes of non-stationary spin-polarized
currents in the stationary state.

For typical $\Gamma\sim1\div10$ meV and $|\varepsilon|\sim50$ meV
\cite{Amaha},\cite{Fransson}, corresponding to the situation
depicted in Fig.\ref{figure3}, Fig.\ref{figure4} the non-stationary
spin-polarized current value is about $1\div10$ nA
($1nA\simeq6\times10^{9}e/sec$).

\begin{figure}
\includegraphics[width=70mm]{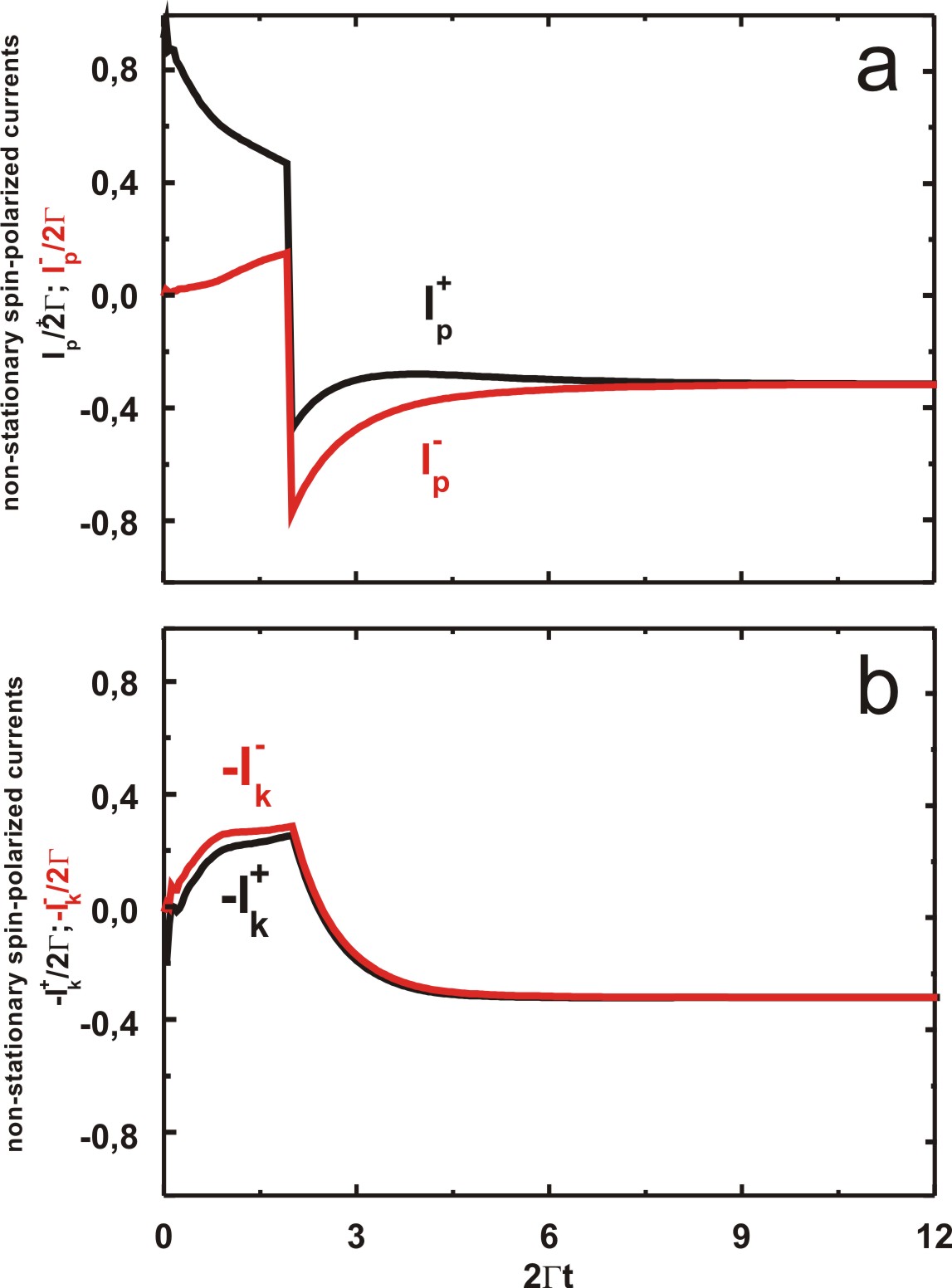}%
\caption{(Color online) Triggering of the normalized non-stationary
spin-polarized tunneling currents in the case, when the value of the
applied bias voltage changes at $2\Gamma t_0=2$. a) Black line
$I_{p}^{+}(t)$, red line $I_{p}^{-}(t)$; b) Black line
$I_{k}^{+}(t)$, red line $I_{k}^{-}(t)$. Parameters $U/2\Gamma=7.5$,
$\varepsilon/2\Gamma=-2.5$, $eV/2\Gamma=-7.5$ for $2\Gamma t<2\Gamma
t_0$ and $eV/2\Gamma=7.5$ for $2\Gamma t>2\Gamma t_0$. Parameter
$\Gamma=1$ is the same for all the figures. $n_{1\sigma}(0)=1$,
$n_{1-\sigma}(0)=0$.} \label{figure6}
\end{figure}

\begin{figure}
\includegraphics[width=70mm]{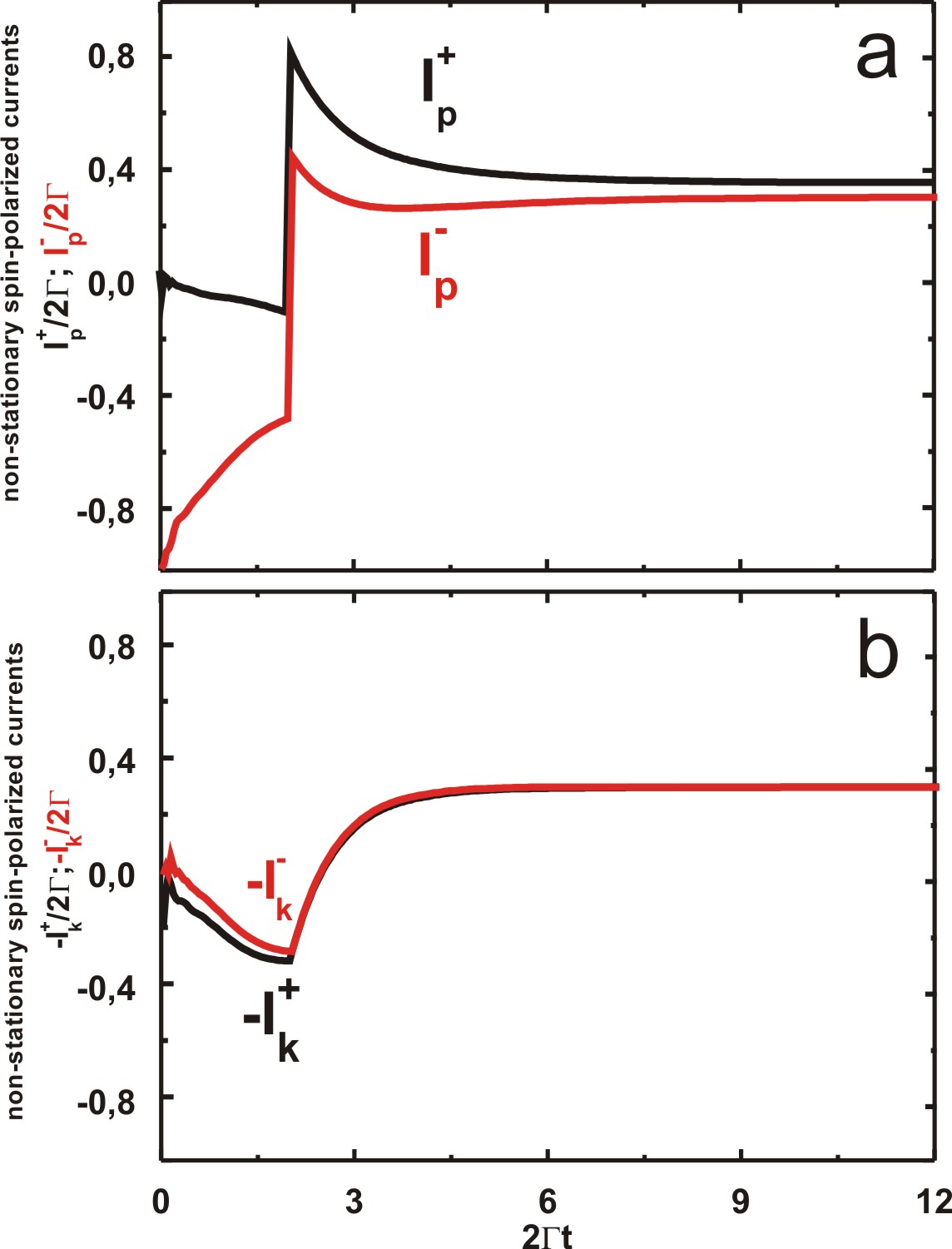}%
\caption{(Color online) Triggering of the normalized non-stationary
spin-polarized tunneling currents in the case, when the value of the
applied bias voltage changes at $2\Gamma t_0=2$. a) Black line
$I_{p}^{+}(t)$, red line $I_{p}^{-}(t)$; b) Black line
$I_{k}^{+}(t)$, red line $I_{k}^{-}(t)$. Parameters $U/2\Gamma=7.5$,
$\varepsilon/2\Gamma=-2.5$, $eV/2\Gamma=7.5$ for $2\Gamma t<2\Gamma
t_0$ and $eV/2\Gamma=-7.5$ for $2\Gamma t>2\Gamma t_0$. Parameter
$\Gamma=1$ is the same for all the figures. $n_{1\sigma}(0)=1$,
$n_{1-\sigma}(0)=0$.} \label{figure7}
\end{figure}

We revealed, that spin polarization and direction of the
non-stationary currents in each lead can be simultaneously inverted
by the sudden changing of the applied bias voltage (see
Fig.\ref{figure6} and Fig.\ref{figure7}). Fig.\ref{figure6}a
demonstrates that initially spin-polarized non-stationary current
with the dominant $I_{p}^{+}(t)$ component changes direction and
polarization (component $I_{p}^{-}(t)$ starts to prevail), when the
applied bias changes the value from the large negative to the large
positive one (system energy scheme changes from the one shown in
Fig.\ref{figure5}a to the one demonstrated in Fig.\ref{figure5}b).
Tunneling current in the another contact lead also changes
polarization and direction (see Fig.\ref{figure6}b), but the
difference between the components with different spins is not so
well pronounced.

Opposite situation is depicted in Fig.\ref{figure7}. In this case
applied bias sign changing leads to the situation when initially
spin-polarized non-stationary current with the dominant component
$I_{p}^{-}(t)$ changes direction and polarization to the opposite
one and component $I_{p}^{+}(t)$ becomes the leading one (see
Fig.\ref{figure6}a) (system energy scheme changes from the one shown
in Fig.\ref{figure5}b to the one demonstrated in
Fig.\ref{figure5}a).

\begin{figure}
\includegraphics[width=70mm]{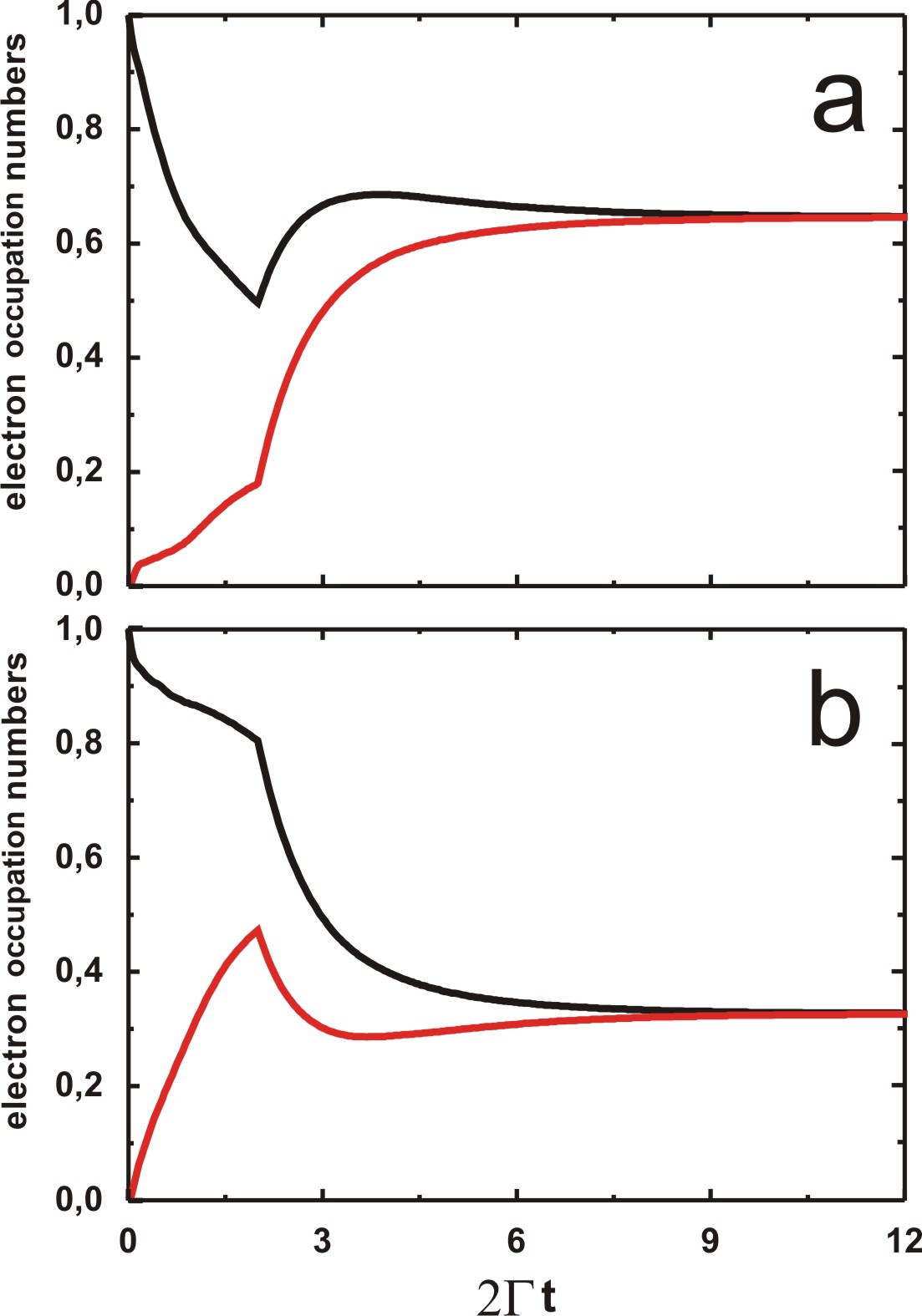}%
\caption{(Color online) Electron occupation numbers time evolution
in the case, when the value of applied bias voltage changes at
$2\Gamma t_0=2$. Black line $n_{1\sigma}(t)$, red line
$n_{1-\sigma}(t)$; a). $eV/2\Gamma=-7.5$ for $t<t_0$ and
$eV/2\Gamma=7.5$ for $2\Gamma t>2\Gamma t_0$ b). $eV/2\Gamma=7.5$
for $t<t_0$ and $eV/2\Gamma=-7.5$ for $2\Gamma t>2\Gamma t_0$.
Parameters $U/2\Gamma=7.5$, $\varepsilon/2\Gamma=-2.5$ and
$\Gamma=1$ are the same for all the figures. $n_{1\sigma}(0)=1$,
$n_{1-\sigma}(0)=0$.} \label{figure8}
\end{figure}

Corresponding electron occupation numbers behavior is shown in
Fig.\ref{figure8}. Electron occupation numbers reveal non-monotonic
behavior.

\section{Stationary correlation functions: formalism and results}

The behavior of the local $"$magnetic$"$ moments can be also
analyzed from the time dependence of the stationary correlation
functions for the electron occupation numbers:

\begin{eqnarray}
K^{\sigma\sigma^{'}}(t-t^{'})=<n_{1\sigma}(t)n_{1\sigma^{'}}(t^{'})>.\nonumber\\
\end{eqnarray}

Correlation functions $K^{\sigma\sigma^{'}}(\tau=t-t^{'})$ satisfy
the system of equations:

\begin{eqnarray}
\frac{\partial}{\partial t}K^{+-}=-2\Gamma\cdot[K^{+-}+\Delta
N^{T}K^{--}-N_{\varepsilon}^{T}n_{1-\sigma}],\nonumber\\
\frac{\partial}{\partial t}K^{--}=-2\Gamma\cdot[K^{--}+\Delta
N^{T}K^{+-}-N_{\varepsilon}^{T}n_{1-\sigma}].
\label{9}\end{eqnarray}

Initial conditions are determined as:

\begin{eqnarray}
K^{+-}(t,t)=K^{+-}(0)=\frac{N_{\varepsilon+U}^{T}\cdot N_{\varepsilon}^{T}}{1+\Delta N^{T}},\nonumber\\
K^{--}(0)=n_{1}^{st}=\frac{N_{\varepsilon}^{T}}{1+\Delta N^{T}}.
\end{eqnarray}

Time evolution of the correlation functions can be obtained from Eq.
(\ref{9}):

\begin{eqnarray}
K^{+-}(\tau=t-t^{'})=\frac{(N_{\varepsilon}^{T})^{2}}{(1+\Delta
N^{T})^{2}}(1-e^{\lambda_{2}^{T}\tau})+\nonumber\\+\frac{N_{\varepsilon}^{T}(N_{\varepsilon+U}^{T}-1)}{2(1+\Delta
N^{T})}e^{\lambda_{1}^{T}\tau}+\frac{N_{\varepsilon}^{T}(N_{k\varepsilon+U}^{T}+1)}{2(1+\Delta
N^{T})}e^{\lambda_{2}^{T}\tau},\nonumber\\
K^{--}(\tau=t-t^{'})=\frac{(N_{\varepsilon}^{T})^{2}}{(1+\Delta
N^{T})^{2}}(1-e^{\lambda_{2}^{T}\tau})+\nonumber\\+\frac{N_{\varepsilon}^{T}(N_{\varepsilon+U}^{T}+1)}{2(1+\Delta
N^{T})}e^{\lambda_{2}^{T}\tau}+\frac{N_{\varepsilon}^{T}(1-N_{\varepsilon+U}^{T})}{2(1+\Delta
N^{T})}e^{\lambda_{1}^{T}\tau}.\nonumber\\
\end{eqnarray}

\begin{figure}
\includegraphics[width=80mm]{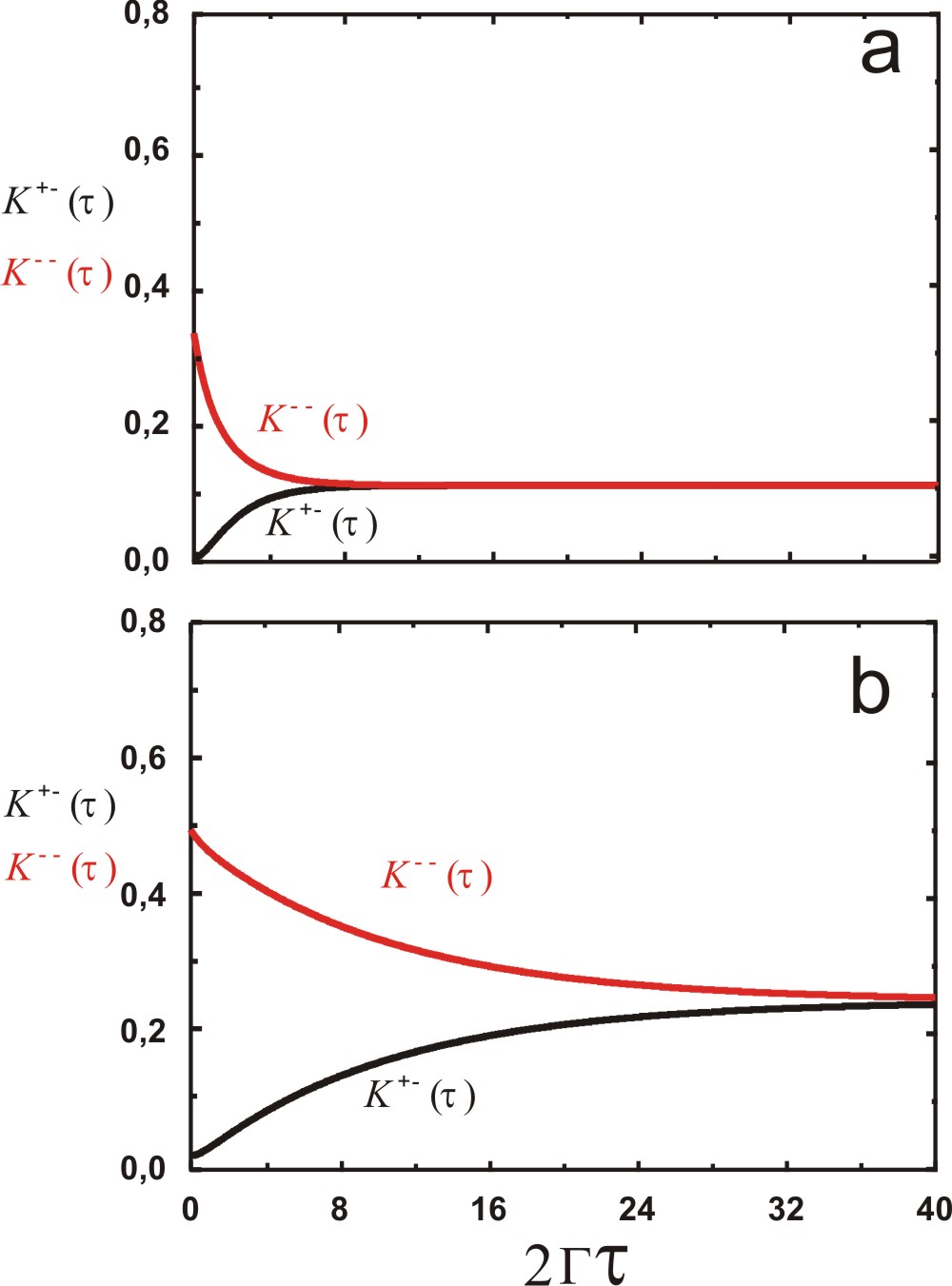}%
\caption{(Color online) Correlation functions time evolution for the
different initial conditions. Black lines demonstrate
$K^{+-}_{\tau}$, red line - $K^{--}_{\tau}$. a) $eV/2\Gamma=-5$; b)
$eV/2\Gamma=2.5$. parameters $U/2\Gamma=7.5$,
$\varepsilon/2\Gamma=-2.5$ and $\Gamma=1$ are the same for all the
figures.} \label{figure9}
\end{figure}

The behavior of the stationary correlation functions for the
localized electrons occupation numbers with the different spin
orientation is depicted in Fig.\ref{figure9}. It is clearly evident,
that for the deep energy levels correlation functions time evolution
is much slower, than for the states with the shallow energy levels.

For $\tau\rightarrow\infty$ correlation functions turn to the
product of the decoupled electronic occupation numbers mean values:

\begin{eqnarray}
K^{+-st}=K^{--st}\simeq(\frac{N_{\varepsilon}^{T}}{1+\Delta
N^{T}})^{2} .
\end{eqnarray}

So for $t<\frac{1}{|\lambda_{1}^{T}|}$ the $"$magnetic$"$
correlations are still present in the system.

\section{Conclusion}
We analyzed time evolution of the opposite spin electron occupation
for the single-localized state with the Coulomb interaction coupled
to two reservoirs in the presence of applied bias voltage. We
revealed that in the presence of the second reservoir with non-zero
applied bias, $"$magnetic$"$ state can be distinguished from the
$"$paramagnetic$"$ one by analyzing time evolution of the electron
occupation numbers. Typical time scales strongly depend on the value
of applied bias and initial conditions.

We revealed that non-stationary spin-polarized currents can flow in
the both leads and their direction and polarization depend on the
value of applied bias. We revealed, that spin polarization and the
direction of the non-stationary currents in each lead can be
simultaneously inverted by the sudden changing of the applied bias
voltage. But in the stationary state occupation numbers for the
electrons with the opposite spins have the same values. Spin
polarized tunneling currents in each lead also become equal.

We also investigated the changes of the time evolution regimes when
the second lead is switched on at the particular time moment. We
found out that switching on of the second lead with the non-zero
applied bias destroys long-living $"$magnetic$"$ moment.

This work was supported by RFBR grant $16-32-60024$ $mol-a-dk$ and
by RFBR grant $14-02-00434$.

 \pagebreak


\begin{thebibliography}{99}
\bibitem{Awschalom}
{\em Semiconductor Spintronics and Quantum Computation}, edited by
D.D. Awschalom, D. Loss, N. Samarth, Nanoscience and Technology
(Springer, Berlin, 2002).
\bibitem{Tsymbal}
E.Y. Tsymbal, O. Mryasov, P.R. LeClair, {\em J. Phys.: Condens.
Matter} {\bf 15}, R109, (2003)
\bibitem{Zutic}
I. \v{Z}uti\'{c}, J. Fabian, S. Das Sarma, {\em Rev. Mod. Phys.}
{\bf 76}, 323, (2004)
\bibitem{Zhu}
H.J. Zhu, M. Ramsteiner, H. Kostial, M. Wassermeier, H.-P.
Schonherr, K.H. Ploog, {\em Phys. Rev. Lett.} {\bf 87}, 116601,
(2001)
\bibitem{Ohno}
Y. Ohno, D.K. Young, B. Beschoten, F. Matsukura, H. Ohno, D.D.
Awschalom, {\em Nature}(London) {\bf 402}, 790, (1999)
\bibitem{Fiederling}
R. Fiederling, M. Keim, G. Reuscher, W. Ossau, G. Schmidt, A. Waag,
L.W. Molenkamp, {\em Nature}(London) {\bf 402}, 787, (1999)
\bibitem{Heersche}
H.B. Heersche, Th. Schapers, J. Nitta, H. Takayanagi, {\em Phys.
Rev. B} {\bf 64}, 161307, (2001)
\bibitem{Egues}
J.C. Egues, {\em Phys. Rev. Lett.} {\bf 80}, 4578, (1998)
\bibitem{Perel'}
V.I. Perel', S.A. Tarasenko, I.N. Yassievich, S.D. Ganichev, V.V.
Bel'kov, W. Prettl, {\em Phys. Rev. B} {\bf 67}, 201304, (2003)
\bibitem{Glazov}
M.M. Glazov, P.S. Alekseev, M.A. Odnoblyudov, V.M. Chistyakov, S.A.
Tarasenko, I.N. Yassievich, {\em Phys. Rev. B} {\bf 71}, 155313,
(2005)
\bibitem{Koga}
T. Koga, J. Nitta, H. Takayanagi, S. Datta, {\em Phys. Rev. Lett.}
{\bf 88}, 126601, (2002)
\bibitem{Voskoboynikov}
A. Voskoboynikov, S.S. Liu, C.P. Lee, {\em Phys. Rev. B} {\bf 58},
15397, (1998)
\bibitem{Bar-Joseph}
I. Bar-Joseph, S.A. Gurvitz, {\it Phys.Rev B}, \textbf{44}, 3332,
(1991).
\bibitem{Gurvitz_1}
S.A. Gurvitz, M.S. Marinov, {\it Phys.Rev A}, \textbf{40}, 2166,
(1989).
\bibitem{Arseyev_1}
P.I. Arseyev, N.S. Maslova, V.N. Mantsevich, {\it European Physical
Journal B}, \textbf{85}(7), 249, (2012).
\bibitem{Stafford_1}
C.A. Stafford, N.S. Wingreen, {\it Phys. Rev. Lett.}, \textbf{76},
1916, (1996).
\bibitem{Hazelzet}
B.L. Hazelzet, M.R. Wegewijs, T. H. Stoof, Y.V. Nazarov, {\it Phys.
Rev. B}, \textbf{63}, 165313, (2001).
\bibitem{Cota}
E. Cota, R. Aguado, G. Platero, {\it Phys. Rev. Lett.}, \textbf{94},
107202, (2005).
\bibitem{Arseyev_2}
P.I. Arseyev, N.S. Maslova, V.N. Mantsevich, {\it Solid State
Comm.}, \textbf{152}, 1545, (2012).
\bibitem{Arseyev_3}
V.N. Mantsevich, N.S. Maslova, P.I. Arseyev, {\it JETP},
\textbf{118}(1), 136, (2014).
\bibitem{Contreras-Pulido}
 L.D. Contreras-Pulido,
J. Splettstoesser, M. Governale, J. Konig, M. Buttiker, {\it Phys.
Rev. B}, \textbf{85}, 075301, (2012).
\bibitem{Elste}
Florian Elste, David R. Reichman, and Andrew J. Millis, {\it Phys.
Rev. B}, \textbf{81}, 205413, (2010).
\bibitem{Kennes}
D. M. Kennes, S. G. Jakobs, C. Karrasch, V. Meden, {\it Phys. Rev.
B}, \textbf{85}, 085113, (2012).

\bibitem{Amaha}
S. Amaha, W. Izumida, T. Hatano, S. Teraoka, S. Tarucha, J. A.
Gupta, and D. G. Austing, {\it Phys. Rev. Lett.}, \textbf{110},
(2013), 016803.
\bibitem{Fransson}
J. Fransson {\it Phys. Rev. B}, \textbf{69}, 201304, (2004).










\end{thebibliography}
\end{document}